\newcommand{\beq}{\begin{equation}}
\newcommand{\eeq}{\end{equation}}
\newcommand{\beqn}{\begin{eqnarray}}
\newcommand{\eeqn}{\end{eqnarray}}

\documentstyle[12pt]{article}
\begin{document}
\begin{titlepage}
\setcounter{page}{0}
\rightline{TECHNION-PH-95-4}

\vspace{2cm}
\begin{center}
{\Large LONG DISTANCE EFFECTS AND CP VIOLATION in $B^{\pm}\to\rho^{\pm}\gamma$}
\vspace{1cm}

{\large G. Eilam$^a$, A. Ioannissian$^{a,b}$, R. R. Mendel$^{a,c}$} \\
\vspace{1cm}
{\em a) Dept. of Physics, Technion - Israel Inst. of Tech., Haifa 32000,
Israel}\\
{\em b) On leave from Yerevan Physics Institute, Alikhanyan Br. 2, Yerevan,
375036, Armenia}\\
{\em c) On sabbatical leave from Dept. of Applied Math., Univ. of Western
Ontario, London, Ontario, Canada}\\
\end{center}

\vspace{5mm}
\centerline{{\bf{Abstract}}}

We demonstrate that the long distance contribution to
$B^{\pm} \to \rho^{\pm} \gamma$
may have a large effect on the decay rate and through it on the CP
violating partial rate asymmetry. The change in
the asymmetry can be between a factor of 0.4 and 1.4 with respect to the
effect of the short distance contribution alone, depending on the values of
the CKM parameters $\rho$ and $\eta$ .

\vfill

\end{titlepage}

\newpage

$1.Introduction$

Interest in rare B decays lies mainly in their potential
role as precision tests of the Standard Model (SM) .
Measurement of the relevant decay rates will provide useful
information about the Cabibbo-Kobayashi-Maskawa (CKM) \cite{ckm}
matrix elements. In particular $|V_{td}|$ is a parameter of crucial
importance in the SM and it is still very poorly known.
It is unrealistic to expect a measurement of $|V_{td}|$ from
$t$-quark decays in the near future.
One would then like to find alternative processes to determine
$|V_{td}|$, e.g. processes that are dominated by virtual
$t$-quarks. The present constraints on $|V_{td}|$
are derived from the experimental value of the
$B_d^o - \bar{B_d^o}$ mixing parameter $x_d \equiv \Delta M / \Gamma$.
The mass difference $\Delta M$ is calculated from the usual box
diagrams. Even though the value of the $t$ quark mass is known \cite{t},
this calculation is still hindered by the uncertainty in the
decay constant $f_B$.
$|V_{td}|$ can be determined precisely from the decay
$K^+ \rightarrow \pi ^+ \nu \bar{\nu} $, which is however very rare.

As is well known, short distance ( SD or penguin) contributions
play a leading role in $b\rightarrow s(d) \gamma$
decays, whose amplitudes are proportional to $V_{ts}$ and
$V_{td}$ respectively.

Since the decays $B\rightarrow K^* \gamma$
have already been observed it is useful to understand what we
may learn about CKM elements through a measurement of
$B\rightarrow \rho \gamma$ decays.

Recent investigation \cite{desh} of the long distance
(LD) quark level spectator contributions to these decays via vector meson
dominance ( VMD )
($b \rightarrow s(d) V \rightarrow s(d) \gamma$)  have shown that
they can be neglected. LD contributions to the decay
$b\rightarrow s(d) \gamma$ via $c\overline{c}$ states
are only a few percent relative to the penguin contribution because
of a strong $q^2$ dependence of the $\psi$  decay constants
( $< 0 | \overline{c} \gamma ^{\mu} c | \psi > = i
\epsilon^{\mu}_{\psi} g_{\psi}(q^2)$ )
$g_{\psi}(m_{\psi}^2) >> g_{\psi}(0)$ .
The LD contributions to the decay $b \rightarrow d \gamma$
via $\rho$ and $\omega$ are only about
$3 \%\cdot  (V_{ub}^{*}/V_{td})$ relative to the penguin
contributions \cite{desh}.

Another source of LD contributions to $b \rightarrow d \gamma$
could come from penguins when we take into account the difference
between $u$ and $c$ quark masses. However Soares \cite{s2} has
claimed that this contribution is only about
$7 \%\cdot(V_{ub}^{*}/V_{td})$
of the penguin, so we will neglect it.
On the other hand LD contributions to the
$B^{\pm}\rightarrow \rho^{\pm} \gamma$
decays that are not of the spectator type
may be significant \cite{asb}.

In this paper we will investigate
$B^{\pm}\rightarrow \rho^{\pm} \gamma$
decays, including the $W$ exchange diagram (ED) in our
analysis (see Fig.1a).
We will show that the ED amplitude is about
$\pm 60 \%\cdot V_{ub}^{*}/ V_{td}$
of the penguin (SD) amplitude (see Fig.1b). The decay rate for
$B^{\pm}\rightarrow \rho^{\pm} \gamma$
can vary by a factor of $0.7 - 2.5$ ( with respect to the
penguin contribution alone ) depending on the CKM matrix elements
$V_{ub}$, $V_{td}$ and on the relative sign of the $W$ exchange
and penguin amplitudes.

We remark that ED and other non-penguin contributions to
$B \to \rho \gamma$ were also studied in some detail in refs.
\cite{cheng,milana}.
However, the approach of ref.\cite{cheng}
suffers from significant double counting in that it includes ED
and VMD contributions additively. It is claimed wrongly that ED and
VMD contributions emanate from two different quark diagrams.
The formalism of ref.\cite{milana}
basically coincides with ours regarding the ED but differs
essentially from our treatment of penguin and related
contributions. For instance the gluons that appear in their
lowest order penguin diagrams are absorbed in our case in meson
wave functions.

Measurement of CP violation in B meson systems would complement
existing results in K decays and help elucidate the origin of this
phenomenon. Observation of the direct CP violation in B decays would
eliminate the superweak model for CP violation.

The main focus of our paper will be the CP rate asymmetry in the
exclusive decays $B^{\pm} \to \rho^{\pm} \gamma$:

\begin{equation}
\label{cp}
a_{cp} \equiv \frac
{\Gamma (B^+ \rightarrow \rho^+ \gamma ) -
\Gamma (B^- \rightarrow \rho^- \gamma ) }
{\Gamma (B^+ \rightarrow \rho^+ \gamma ) +
\Gamma (B^- \rightarrow \rho^- \gamma ) }.
\end{equation}
which was investigated in
ref.
\cite{w2}
without taking into account ED contributions.

An important finding of our investigation
is that the exchange diagram does not contribute significantly
to the rate difference of the decays of charge conjugate channels
$B^{\pm}$ (the numerator of eq.(\ref{cp})).
As was mentioned above, the rate ( i.e. the  denominator of
eq.(\ref{cp})), can change by a factor $0.7-2.5$.
So the CP asymmetry for the decays
$B^{\pm} \rightarrow \rho^{\pm} \gamma$ will change by the same
factor as the rates (compared to the SD contribution by itself ).
The resulting CP asymmetry could be as large as $30 \%$ within the allowed
range of CKM matrix elements.

Because the ED contribution is negligible for the
processes $B^{\pm} \to K^{* \pm} \gamma$ due to the relevant CKM
factors, it can also importantly affect the interesting ratio

\beq
\label{r}
R \equiv \frac{
\Gamma(B^+\to \rho^+ \gamma)+
\Gamma(B^-\to \rho^-\gamma) }
{\Gamma(B^+\to K^{*+} \gamma)+
\Gamma(B^-\to K^{*-} \gamma) }
\eeq

\vspace{1cm}

$2.W$ $Exchange$ $Diagram$ $contribution$ $to$
 $B^{\pm} \to \rho^{\pm} \gamma $

It is well known that in some processes ED can be appreciably large
i.e. of the same order or even larger than the SD (penguin)
 see Fig.1b,  contribution \cite{gad,rob}. The exclusive decays
$ B^{\pm}\rightarrow \rho^{\pm} \gamma $
, can occur via ED as shown in Fig.1a. In a simple constituent quark model
for the B and $\rho$ mesons ( and neglecting terms of order
$M_{\rho}^2/M_B^2$), the amplitude is given by \cite{asb,rob}

\beqn
\label{e1}
A_{ED}( B^- \rightarrow \rho^- \gamma ) &=& \frac{\sqrt{2} G_F f_B f_{\rho}
M_{\rho}}{6 M_B m_u}  e V_{ub} V_{ud}^* \nonumber\\
                                        & & Tr[( \not{P_B} -
M_B ) \not{\varepsilon_{\gamma}} \not{q_{\gamma}} \not{\varepsilon_{\rho}}
( \frac{ 1-\gamma_5 }{2})]
\eeqn
The QCD correction factor $[C_2(M_B)+C_1(M_B)/3] \approx 1$
has been absorbed, $m_u$ = 330 MeV is the constituent mass of the
$u$ quark and we used the following definition for the decay constants
$f_B$ and $f_{\rho}$ ( $f_{\rho} \approx 216MeV$ )

\begin{equation}
\label{frfb}
< 0 | \overline{u} \gamma_{\mu}  d | \rho^- > \equiv i
\varepsilon^{\mu}_{\rho} M_{\rho} f_{\rho} ,\;\;\;\;\;\;\;
< 0 | \overline{u} \gamma_{\mu} \gamma_5 b | B_u^- > \equiv i f_{B}
P_B^{\mu}.
\end{equation}

The factor $m_u$ in the denominator of eq.(\ref{e1}) indicates clearly
that the ED is a long distance contribution. The effective value for
$m_u$ that should be used could easily be $30 \%$ larger
or smaller than the commonly used value quoted above. This then
constitutes an intrinsic uncertainty in our estimate of the ED
amplitude ( see however a comment at the end of the paper ).

Evaluating the trace in (\ref{e1}) we get

\beqn
\label{e2}
A_{ED}( B^- \rightarrow \rho^- \gamma )&=& \frac{\sqrt{2} G_F f_B f_{\rho}
M_{\rho}}{3 M_B m_u}  e V_{ub}
V_{ud}^*  \nonumber\\
                                  & &\{ i \epsilon_{\mu \nu \alpha \beta}
\varepsilon_{\gamma}^{\mu} q_{\gamma}^{\nu}
\varepsilon_{\rho}^{\alpha} p_{\rho}^{\beta}
+( \varepsilon_{\gamma} p_{\rho}) ( q_{\gamma} \varepsilon_{\rho})
-( p_{\rho} q_{\gamma}) ( \varepsilon_{\gamma} \varepsilon_{\rho}) \} .
\eeqn
We denote by $p_V$ ($p_{\gamma}$) and
$\varepsilon_V$ ($\varepsilon_{\gamma}$) the momentum and
polarization of the vector meson (photon), respectively.

\vspace{1cm}

$3.Short$ $Distance$ $contribution$ $to$
$B\to \rho \gamma $
$and$
$B\to K^* \gamma $

Now we consider the SD contribution to the $B\to V
\gamma$ decay ($V = \rho, K^*$). We will follow
refs. \cite{w2,w1} for the calculation of the exclusive
matrix elements for $B\to V \gamma$, adding $SU(3)_{flavour}$
breaking effects for the vector meson wave functions \cite{zhit,ali}.

The SD Hamiltonian for the decay $b\to s(d) \gamma$, is

\beq
\label{s1}
H = -\frac{4 G_F}{\sqrt{2}}v_q C_7 O_7,
\eeq
where $q$ = $s$ or $d$  and $v_s=V_{tb} V_{ts}^*$, $v_d=V_{tb} V_{td}^*$.
The magnetic dipole operator $O_7$ is (in the limit $\frac{m_s}{m_b}\to 0$)

\beq
\label{o7}
O_7=\frac{e}{16 \pi^2 } \bar{q}_L \sigma^{\mu \nu } m_b b_R F_{\mu \nu },
\eeq
and we will use $C_7(5\;GeV)=-.305$  \cite{w2}.
The amplitude for the exclusive decay $B\to V \gamma$ is then
\beq
\label{s2}
A_{SD}(B\to V \gamma) = -\frac{4 G_F}{\sqrt{2}}v_q C_7 < V \gamma|O_7|B >
\eeq

The transition matrix element of the magnetic operator $O_7$ for an
on-shell photon has the general form (we neglect the vector meson mass)

\beqn
\label{mo7}
<V \gamma |O_7 |B> =
e M_B F_V(O_7)
\{ i \epsilon_{\mu \nu \alpha \beta}
\varepsilon_{\gamma}^{\mu} q_{\gamma}^{\nu}
\varepsilon_{\rho}^{\alpha} p_{\rho}^{\beta}\nonumber\\
+( \varepsilon_{\gamma} p_{\rho}) ( q_{\gamma} \varepsilon_{\rho})
-( p_{\rho} q_{\gamma}) ( \varepsilon_{\gamma} \varepsilon_{\rho})\}
\eeqn

In the model of ref.\cite{w2}  $F(O_7)$ is given by

\beq
\label{fo7}
F_V(O_7)=-\frac{\sqrt{M_B} C_B C_V}{4 \pi ^2} \int_0^1 d y
\frac{y}{\sqrt{1+y}} \phi_V (y) \phi_B \big( \frac{M_B (1-y)}{2} \big).
\eeq
Here $\phi_B(p)= exp \big( -p^2/(2p_F^2) \big)$ is the harmonic
oscillator wave function. $C_B=\sqrt{8} \pi^{3/4}p_F^{-3/2}$ and
$C_V= f_V/ ( 4 \sqrt{3} )$ are the normalization factors of the
B meson and vector meson wave functions, respectively.
$\phi_V(y)$ is the quark distribution
amplitude in the vector meson ( $p_{s(d)}=p_V y$ and $p_u=p_V(1-y)$).
For the $\rho$ and $K^*$ mesons we take, respectively \cite{ali}

\beq
\label{fr}
\phi(y)_{\rho}=6 y ( 1- y ) \{ 1 - 0.85 [(2 y -1)^2- \frac{1}{5}] \}
\eeq
and

\beqn
\label{fk}
\phi(y)_{K^*}&=&6 y ( 1- y )
\{ 1 + 0.57 (2y-1)- \nonumber\\
             & &1.35 [(2 y -1)^2-
\frac{1}{5}] +0.46 [ \frac{7}{3}(2y-1)^3-(2y-1)] \}.
\eeqn

$4.The$ $decay$ $rate$ for
$B^{\pm}\to \rho^{\pm} \gamma$

Before presenting our results we would like to note
that the relative sign of the ED and penguin amplitudes is not known.
We will assume that ED has no strong phase. Such phase could arise for
example from order $\alpha_s^2$ QCD corrections which however are
expected to be small because $\alpha_s(m_b) \approx 0.2$. Adding
eqs. (\ref{e2}) and (\ref{s2}) we get

\beqn
\label{sl1}
A(B^-\to \rho^- \gamma) = -\frac{4 G_F}{\sqrt{2}} C_7 V_{tb}V_{td}^*
e M_B F_{\rho}(O_7)
\frac{A_{SD}+A_{ED}}{A_{SD}} \nonumber\\
\{ i \epsilon_{\mu \nu \alpha \beta}
\varepsilon_{\gamma}^{\mu} q_{\gamma}^{\nu}
\varepsilon_{\rho}^{\alpha} p_{\rho}^{\beta}
+( \varepsilon_{\gamma} p_{\rho}) ( q_{\gamma} \varepsilon_{\rho})
-( p_{\rho} q_{\gamma}) ( \varepsilon_{\gamma} \varepsilon_{\rho})\}
\eeqn
where

\beq
\label{spe1}
\frac{A_{SD}+A_{ED}}{A_{SD}}=
\{1 \mp \frac{f_B f_{\rho} M_{\rho}}{6 M_B^2 m_u C_7
F_{\rho}(O_7)}\frac{V_{ub}}{V_{td}^*} \}
\eeq

The rate for
$B^-\to \rho^- \gamma$
 is then given by

\beq
\label{gr1}
\Gamma (B^- \to \rho^- \gamma ) = 2 G_F^2 M_B^5 \alpha C_7^2
|F_{\rho}(O_7)|^2 |V_{td}|^2 \Omega
\eeq
where

\beq
\label{omega}
\Omega \equiv \Big| \frac{A_{SD}+A_{ED}}{A_{SD}}\Big|^2.
\eeq

As is seen from eq.(\ref{e2}) the ED contribution is proportional to the
B meson decay constant $f_B$. The  $B_d^o-\bar{B_d^o}$ mixing parameter
$x_d$ is proportional to $f_B^2$ \cite{ros}:

\beq
\label{deltam}
x_d = \tau_B \frac{G_F^2}{6 \pi^2} |V_{td}|^2 m_t^2 M_B f_B^2 B_B
\eta_{QCD} F(m_t^2/M_W^2)
\eeq
where $\eta_{QCD}$ is a QCD correction factor and

\beq
\label{f(x)}
F(x)= \frac{1}{4} \Big[ 1+ \frac{3-9x}{(x-1)^2}+\frac{6x^2 \ln
x}{(x-1)^3}\Big]
\eeq
Here $x=m_t^2/M_W^2$ and

\beq
\label{mm}
m_t=\frac{m_t^{pole}}{ [1+\frac{4 \alpha_s (m_t)}{3
\pi}+K_t (\frac{\alpha_s (m_t)}{\pi})^2] }
\eeq
where $K_t \simeq 11$ \cite{mik}.

Substituting $f_B$ from eq.(\ref{deltam}) in eq.(\ref{spe1}) we get

\beq
\label{spe2}
\frac{A_{SD}+A_{ED}}{A_{SD}}=
\{1 \mp \frac{f_{\rho} M_{\rho}}{6 M_B^2 m_u C_7
F_{\rho}(O_7)}
\frac{V_{ub}}{|V_{td}| V_{td}^*}
\sqrt{\frac{x_d 6 \pi^2}{\tau_B G_F^2 m_t^2 M_B B_B \eta_{QCD}
F(m_t^2/M_W^2)}} \}
\eeq

Using the Wolfenstein parametrization for the CKM matrix elements \cite{wol}
 we obtain

\beq
\label{spe3}
\frac{A_{SD}+A_{ED}}{A_{SD}} =
\{1 + g
\frac{\rho-\rho^2-\eta^2-i \eta}{[(1-\rho)^2+\eta^2]^{3/2}}
\}
\eeq
where $g$ is given by

\beq
\label{g}
g =  \mp \frac{f_{\rho} M_{\rho}}{6 M_B^2 m_u C_7
F_{\rho}(O_7)}
\sqrt{\frac{x_d 6 \pi^2}{\tau_B G_F^2 m_t^2 M_B B_B \eta_{QCD}
F(m_t^2/M_W^2)}} \frac{1}{A \lambda^3}
\eeq

For our estimates we use the following input parameters \cite{ros}:
$x_d=0.67 \pm 0.10$ , $\tau_B=(1.49 \pm 0.04) 10^{-12} s $,
$m_t^{pole}=176$ GeV \cite{t} ( $m_t=167$
GeV ) , $\eta_{QCD}=0.85 \pm 0.05$,  $B_B=1$ , $\lambda =0.22$, $A=0.785
\pm 0.093$, $m_u=330 MeV$. For the central values of these parameters
we get $|g|=0.6$ .

For $\Omega$ in the Wolfenstein parametrization of the CKM matrix
we have

\beqn
\label{omeg}
\Omega(\rho, \eta) &\equiv& \Big| \frac{A_{SD}+A_{ED}}{A_{SD}}\Big|^2
\nonumber\\
                   &=     &  \{1 + g
\frac{\rho-\rho^2-\eta^2}
{[(1-\rho)^2+\eta^2]^{3/2}}
\}^2 + \frac{g^2 \eta^2}{
[(1-\rho)^2+\eta^2]
^3}
\eeqn

In Fig.2 we plot curves corresponding to constant
values of $\Omega$ for $g=0.6$ (Fig.2a) and $g=-0.6$ (Fig.2b).
Note that if the ED is not taken into account ( $g=0 $) , then
$\Omega=1$ for all $\rho , \eta$ .

We indicate in the Figs.2-5
 the region in
the $\rho , \eta$ plane that is allowed by present experimental bounds.
The constraints on $\rho$ and $\eta$ are obtained
 \cite{ros} from the
following experimental results

\beqn
\label{constr}
\bigg|\frac{V_{ub}}{V_{cb}}\bigg|
= 0.08 \pm 0.03 \nonumber\\
|\epsilon_K| = (2.26 \pm 0.02) \cdot 10^{-3}
\eeqn

The bounds on $\bigg|\frac{V_{ub}}{V_{cb}}\bigg|$ give
$ \sqrt{\rho^2+\eta^2}=\frac{1}{.22}(0.08 \pm 0.03)$ which leads to two
circles centered at (0,0).

The bounds on $\epsilon_K$ lead to the hyperbolas \cite{ros}

\beq
\label{eta}
\frac{1}{\eta}=1.5\cdot A^2 B_K+6.6\cdot A^4 B_K\cdot (1-\rho)
\eeq
where we will take $A=0.785 \pm 0.093$, $B_K=0.8\pm0.2$ \cite{ros}.

For the branching ratio of the decay $B^- \to \rho^- \gamma$ we get

\beqn
\label{br}
Br (B^- \to \rho^- \gamma )&=& 2 \tau_B G_F^2 M_B^5 \alpha C_7^2
|F_{\rho}(O_7)|^2 A^2 \lambda^6
[(1-\rho)^2+\eta^2] \Omega(\rho, \eta) \nonumber\\
                           &\simeq&8.2\cdot10^{-7}[(1-\rho)^2+\eta^2]
\Omega (\rho, \eta)\cdot \nonumber\\
                           & &
\Big(\frac{C_7}{-0.305}\Big)^2
\;\Big(\frac{F_{\rho}(O_7)}{-2.6\cdot10^{-3}}\Big)^2
\;\Big(\frac{A}{0.785}\Big)^2
\eeqn

For the ratio $R$ in the Wolfenstein parametrization we get

\beqn
\label{rr}
R &\equiv& \frac{
\Gamma(B^-\to \rho^- \gamma) +
\Gamma(B^+\to \rho^+ \gamma) }
 {\Gamma(B^-\to K^{* -} \gamma) +
\Gamma(B^+\to K^{* +} \gamma) }
=\bigg| \frac{F_{\rho}(O_7)}{F_{K^*}(O_7)}\bigg|^2
\frac{|V_{td}|^2}{|V_{ts}|^2}\cdot \Omega(\rho, \eta)
\nonumber\\
  &=     &
\bigg| \frac{F_{\rho}(O_7)}{F_{K^*}(O_7)}\bigg|^2 \lambda^2
[(1-\rho)^2+\eta^2] \Omega(\rho, \eta)
\eeqn
where
$| \frac{F_{\rho}(O_7)}{F_{K^*}(O_7)}|^2=0.545
\pm .025$
for $p_F=0.5 - 0.65$ GeV.

As an illustration we plot curves corresponding to constant values
of $R$ for $g=+0.6$ (Fig.3a) and $g=-0.6$ (Fig.3b) in the $\rho ,
\eta$ plane.

\vspace{1cm}

$5.CP$ $asymmetry$ $in$ $the$ $decay$
$B^{\pm} \to \rho^{\pm} \gamma$

We now consider direct CP violation in the decay
$B^{\pm} \to \rho^{\pm} \gamma$.

In the spectator approximation the absorptive contribution to the quark
decay $b\to d \gamma$ comes from the real intermediate states
$b \to u \bar{u} d \to d \gamma$,
$b \to c \bar{c} d \to d \gamma$,
$b \to g d \to d \gamma$. The absorptive part of Hamiltonian in the
spectator approximation is \cite{s1}

\beq
\label{im}
H_{ab}\simeq -i \alpha_s \frac{4G_F}{\sqrt{2}}O_7(
\frac{1}{4}C_2 \cdot V_{ub}V_{ud}^*+
0.12\frac{1}{4}C_2 \cdot V_{cb}V_{cd}^*+
\frac{2}{9} C_8 \cdot V_{tb}V_{td}^*)
\eeq
Here $C_2(5\;GeV)=1.096$ \cite{w2} and $C_8(5\;GeV)=-0.185$ \cite{mis}.

As explained earlier, the
absorptive part of the ED amplitude is of order $\alpha_s^2$. Moreover
the interference of the ED amplitude with the first term in
$H_{ab}$ will not contribute to the numerator of eq. (\ref{cp}) either.
We have also estimated the contribution stemming from the interference of
the ED amplitude with the last two terms in eq.(\ref{im}) and found
that it enhances (reduces) for $g=-0.6$ ($g$=+0.6) the numerator of
$a_{cp}$ by approximately $12 \%$ with respect to
the leading contribution alone. We will therefore ignore it together with
the order $\alpha_s^2$ contributions. We conclude that
the ED can affect significantly only the
denominator of $a_{cp}$ in eq. (\ref{cp}).
To estimate the numerator of the asymmetry in eq.(\ref{cp}) we follow
the formalism of \cite{w2} which includes two types of contributions. The
first results from the interference of the first two terms in $H_{ab}$
with the leading penguin amplitude ( see eqs.(\ref{s1})-(\ref{mo7})) while
the second includes the interference of the absorptive parts in
non-spectator diagrams (see Fig. 3 in ref. \cite{w2}) with the leading
penguin amplitude.
Since the absorptive part comes from long distance effects, a cutoff
parameter $\Lambda$ for the gluon momentum, is introduced to avoid double
counting with the contribution from the wave functions of the mesons
\cite{w2}.
The denominator of eq.(\ref{cp}) is calculated using eq.(\ref{gr1})
($\Gamma(B^- \to \rho^- \gamma)+\Gamma(B^+\to \rho^+ \gamma)\simeq
2\Gamma(B^- \to \rho^- \gamma)$).

The asymmetry $a_{cp}$ is then given by

\beq
\label{acp}
a_{cp}= -K_{\Lambda}\frac{\eta}{[(1-\rho)^2+\eta^2]}
\frac{\alpha_s}{2}
\frac{C_2}{C_7}
\frac{1}{\Omega (\rho, \eta)}
\eeq
where $K_{\Lambda} \simeq 0.97$ for $\Lambda=1$ GeV
and $K_{\Lambda} \simeq 1.3$ for $\Lambda=0.2$ GeV \cite{w2}.

In Fig.4 we plot curves of constant $a_{cp}$ ( a few representative
values) for $g=+0.6$ (Fig.4a) and $g=-0.6$ (Fig.4b) for cut-off
parameter $\Lambda=0.2$ GeV and $\alpha_s=0.2$.

We now estimate the total number of charged B decays required to observe
this CP asymmetry at the $3\sigma$ level, $N_{3\sigma}$, assuming
perfect experimental detection efficiency

\beqn
\label{n}
N_{3\sigma}&=& \frac{9}{a_{cp}^2 Br(B^- \to \rho^- \gamma) }
\nonumber\\
           &\approx&  5 \cdot 10^7 \;\frac{(1-\rho)^2+\eta^2}{\eta^2}\cdot
\Omega (\rho,\eta)\cdot \\
           & &
\Big(\frac{1.3}{K_{\Lambda}}\Big)^2
\Big(\frac{0.2}{\alpha_s}\Big)^2
\Big(\frac{-2.6\cdot10^{-3}}{F_{\rho}(O_7)}\Big)^2
\Big(\frac{0.785}{A}\Big)^2 \nonumber
\eeqn

In Fig.5 we plot curves of constant $N_{3\sigma}$
for $g=+0.6$ (Fig.5a) and $g=-0.6$ (Fig.5b).

\vspace{1cm}

Summarizing our main results, Fig.4 indicates that there can be a sizable
CP asymmetry in $B^{\pm} \to \rho^{\pm} \gamma$, possibly as large as 30\%.
Comparison of Figs. 4a and 4b shows that the long distance ED amplitude
plays an important role in determining the value of the asymmetry ( a
factor 0.4-1.4 with respect to the pure penguin contribution), and also
underlines the need to theoretically understand the relative sign of the
penguin and ED amplitudes. The importance of the annihilation diagram is
also apparent from Figs 3a and 3b, where the ratio $R$ defined in eq.
(\ref{r}) is shown. The plot of Fig.5
shows that about $10^8-10^9$ charged B mesons are needed to be able to
observe the CP partial rate asymmetry at the $3\sigma$ level under ideal
experimental
conditions. Such numbers are not out of the question for near future
B factories. Finally, let us note that the magnitude of the ED
contribution as presented here will be determined by checking against
experiment the predictions of ref. \cite{18} for $B\to l \nu \gamma$
where the ED is supposed to dominate the amplitude, leading to a
branching ratio of about $4\cdot10^{-6}$ for $m_u=330$ MeV.

\vspace{1cm}

$Acknowledgements$

This research is supported in part by Grant 5421-1-94 from the Ministry
of
Science and the Arts of Israel and by the Jubilee Fund of the Austrian
National Bank, Project 5051.
A.I. has been partially supported by the Lady Davis Trustship.
The work of R.M. was supported in part by the Natural Sciences and
Engineering Research Council of Canada. The work of G.E. has been
supported in part by the Fund for Promotion of Research at the
Technion and by GIF. We would like to thank B.Blok for useful discussions.

\newpage

\begin{center}

Figure Captions

\end{center}

\vspace{2cm}

Fig.1a The $W$ Exchange Diagram (ED) for the decay $B\to\rho\gamma$

\vspace{5mm}

Fig.1b The electromagnetic penguin diagram for the decay $B \to V \gamma$
 ($V=K^*,\rho$).

\vspace{5mm}

Fig.2a
Curves corresponding to constant
$\Omega=0.75,0.85,1.0,1.5,2.0,2.5$ for $g=0.6$.
The full lines show the constraints from $|V_{ub}|/|V_{cb}|$ and
$|\epsilon_K|$, i.e. here and in all the following figures, the
region in $(\rho,\eta)$ within the full lines is allowed by experiment
(see ref.\cite{ros}).

\vspace{5mm}

Fig.2b
Curves corresponding to constant
$\Omega=0.7,0.85,1.0,1.15,1.28$
for $g=-0.6$.

\vspace{5mm}

Fig.3a
Curves corresponding to constant  $R=0.025,0.03,0.035,0.04$
for $g=0.6$ and $| \frac{F_{\rho}(O_7)}{F_{K^*}(O_7)}|^2=0.545$.

\vspace{5mm}

Fig.3b
Curves corresponding to constant  $R=0.01,0.03,0.05,0.07$
for $g=-0.6$ and $| \frac{F_{\rho}(O_7)}{F_{K^*}(O_7)}|^2=0.545$.

\vspace{5mm}

Fig.4a
Curves corresponding to constant
$a_{cp}=5\%,10\%,15\%,20\%$
for $g=0.6$, $\alpha_s=0.2$ and $K_{\Lambda}=1.3$.

\vspace{5mm}

Fig.4b
Curves corresponding to constant
$a_{cp}=5\%,10\%,15\%,20\%,25\%,$ $30\%$
for $g=-0.6$, $\alpha_s=0.2$ and $K_{\Lambda}=1.3$.

\vspace{5mm}

Fig.5a
Curves corresponding to constant
$N_{3\sigma}=2\cdot10^8,5\cdot10^8,10^9,5\cdot10^9$
for $g=0.6$, $\alpha_s=0.2$, $K_{\Lambda}=1.3$, $A=0.785$ and
$F_{\rho}(O_7)=-2.6\cdot10^{-3}$.

\vspace{5mm}

Fig.5b
Curves corresponding to constant
$N_{3\sigma}=2\cdot10^8,5\cdot10^8,10^9,5\cdot10^9$
for $g=-0.6$, $\alpha_s=0.2$, $K_{\Lambda}=1.3$, $A=0.785$ and
$F_{\rho}(O_7)=-2.6\cdot10^{-3}$.

\end{document}